\documentclass[%
 reprint,
 amsmath,amssymb,
 aps,
prb,
 superscriptaddress
]{revtex4-2}

\usepackage{graphicx}% Include figure files
\usepackage{dcolumn}% Align table columns on decimal point
\usepackage{bm}% bold math
\usepackage{color}
\begin{document}

\preprint{APS/123-QED}

\title{Non-Hermitian physics of levitated nanoparticle array}% Force line breaks with \\

\author{Kazuki Yokomizo}
\affiliation{Department of Physics, The University of Tokyo, 7-3-1 Hongo, Bunkyo-ku, Tokyo, 113-0033, Japan}
\author{Yuto Ashida}%
\affiliation{Department of Physics, The University of Tokyo, 7-3-1 Hongo, Bunkyo-ku, Tokyo, 113-0033, Japan}
\affiliation{Institute for Physics of Intelligence, University of Tokyo, 7-3-1 Hongo, Tokyo 113-0033, Japan}


%\affiliation{
%\affiliation{
%\affiliation{%


\begin{abstract}
The ability to control levitated nanoparticles allows one to explore various fields of physics, including quantum optics, quantum metrology, and nonequilibrium physics. It has been recently demonstrated that the arrangement of two levitated nanoparticles naturally realizes the tunable nonreciprocal dipole-dipole interaction. Motivated by this development, we here propose and analyze an array of levitated nanoparticles as an ideal platform to study non-Hermitian physics in a highly controlled manner. We employ the non-Bloch band theory to determine the continuum bands of the proposed setup and investigate the non-Hermitian skin effect therein. In particular, we point out that the levitated nanoparticle array exhibits rich dynamical phases, including the dynamically unstable phase and the unconventional critical phase where the spectral singularity persists over a broad region of the controllable parameters. We also show that the long-range nature of the dipole-dipole interaction gives rise to the unique self-crossing point of the continuum band.
%\begin{description}
%\end{description}
\end{abstract}
\pacs{Valid PACS appear here}% PACS, the Physics and Astronomy
\maketitle
%
%\tableofcontents
%

\section{\label{sec1}Introduction}
A levitated nanoparticle is a laser trapped nanoscale dielectric particle smaller than wavelength of light~\cite{Ashkin1986}. Recent experimental developments have allowed one to cool a levitated nanoparticle to ultracold temperatures~\cite{Gieseler2012,Asenbaum2013,Millen2015,Delic2019,Iwasaki2019,Kamba2021,Kamba2022,Vijayan2023,Piotrowski2023} and offered unique opportunities to study quantum mechanics of mesoscopic objects~\cite{Chang2010,Barker2010,Romero2010,Nimmrichter2010,Romero2011,Romero2011v2,Tebbenjohanns2020,De2021,Delic2020,Magrini2021,Tebbenjohanns2021,Ranfagni2021}. Additionally, previous studies demonstrated the potential of a levitated nanoparticle to explore various fields of physics, such as nonequilibrium physics~\cite{Monteiro2013,Gieseler2013,Gieseler2014,Gieseler2014,Fonseca2016,Ricci2017,Rondin2017,Hoang2018} and quantum sensing~\cite{Arvanitaki2013,Bateman2014,Geraci2015,Ranjit2016,Hempston2017,Hebestreit2018,Reimann2018,Ricci2019,Zheng2020,Weiss2021,Rudolph2022}. Remarkably, recent experimental studies have shown the possibility of realizing multi-particle setups~\cite{Tatarkova2002,Mohanty2004,Guillon2006,Brzobohaty2010,Chang2013,Yan2015,Shahmoon2017,Li2020,Rieser2022,Yu2022,Yan2023}. In particular, Ref.~\cite{Yan2023} has reported a realization of an on-demand assembly of levitated nanoparticles, in which optical tweezers are used to trap and arrange the nanoparticles one by one.

On another front, recent years have witnessed remarkable advances in our understandings of non-Hermitian systems, i.e., a class of nonequilibrium systems that can be effectively described by non-Hermitian operators~\cite{Ashida2020}. While non-Hermitian physics has been widely investigated in several fields of quantum science, such as ultracold atoms~\cite{Li2019,Gou2020,Takasu2020,Ozturk2021,Liang2022} and photonics~\cite{Xiao2020,Weidemann2020,Wang2021,Xiao2021}, its idea has also found numerous applications in classical systems realized in optics~\cite{Guo2009,Feng2013,Zhen2015,Zhou2018}, mechanics~\cite{Brandenbourger2019,Ghatak2020,Chen2021,Wang2022}, and electrical circuits~\cite{Rosenthal2018,Helbig2020,Hofmann2020,Zou2021}. These previous studies uncovered rich non-Hermitian phenomena that have no counterparts to Hermitian systems. For instance, one-dimensional (1D) tight-binding systems with asymmetric hopping amplitudes exhibit the non-Hermitian skin effect~\cite{Yao2018,Okuma2020,Zhang2020}, where the bulk eigenstates are localized at open boundaries, leading to the extreme boundary sensitivity of the eigenvalue.

In this paper, we propose and analyze a 1D levitated nanoparticle array as an ideal platform to study previously unexplored regimes of non-Hermitian physics in a highly controlled manner. A prominent feature is that there exists the tunable nonreciprocal dipole-dipole interaction between particles, which is induced by the nonreciprocal interference originating from phase difference between the trapping lasers~\cite{Rieser2022}. The proposed system then realizes a 1D tight-binding model with arbitrarily tunable asymmetric hopping amplitudes that have possibly negative signs and long-range dependence. This high controllability allows one to explore the whole parameter region of non-Hermitian systems, thus opening the possibility to fully uncover the potential of non-Hermitian systems. The proposed setup should be contrasted to the previous non-Hermitian platforms where it remains challenging to realize long-range asymmetric and/or negative hopping amplitudes.

To determine the continuum bands and the dynamical phase diagram of the levitated nanoparticle array, we invoke the non-Bloch band theory~\cite{Yao2018,Yokomizo2019,Kawabata2020,Yokomizo2020,Yokomizo2021,Yokomizo2022}, a recently developed powerful tool to investigate models featuring the non-Hermitian skin effect. The non-Bloch band theory allows for calculating the asymptotic eigenvalues under open boundary conditions in the limit of a large system size. This makes contrast to the conventional Bloch band theory, where the band structure reproduces the eigenvalues under periodic boundary conditions.

On the basis of this theoretical framework, we find that the levitated nanoparticle array exhibits rich dynamical phases, including the unconventional critical phase and the dynamically unstable phase. In the former, a remarkable feature is that the non-Hermitian degeneracy of the continuum bands known as the spectral singularity appears over a broad region of the parameters. The key ingredients of the latter are negative interparticle couplings, which were difficult to realize in the existing non-Hermitian platforms. Moreover, the proposed system can naturally realize the long-range hopping amplitudes originating from the dipole-dipole interaction. We show that this long-range nature leads to the unique self-crossing point of the continuum band, which corresponds to the singularity of the generalized Brillouin zone.

%
% -------------------------------------------------------------------------------------------------------------------------------------------------------------------------------------
%

\section{\label{sec2}Levitated nanoparticle array}
\begin{figure}[]
\includegraphics[width=8.5cm]{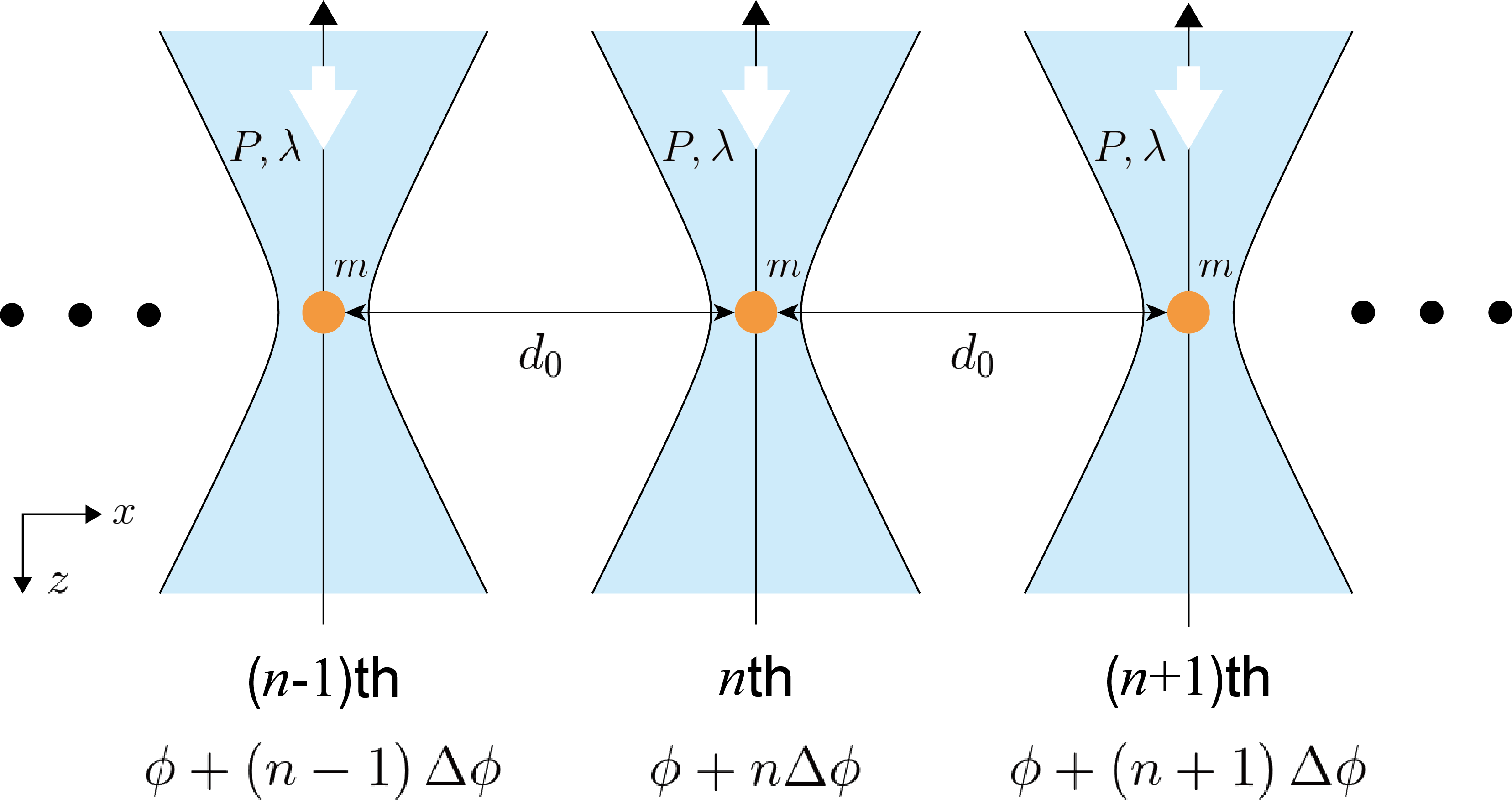}
\caption{\label{fig1}Schematic figure of the levitated nanoparticle array. The distance between the nearest-neighbor particles is $d_0$, and the mass of all the particles is $m$. All the trapping lasers have the power $P$ and the wavelength $\lambda$. We set the phase of the $n$th trapping laser in the focal plane to be $\phi+n\Delta\phi$.}
\end{figure}
To be concrete, we consider a 1D array of the trapped levitated nanoparticles (Fig.~\ref{fig1}). The particles are equally spaced at the interval $d_0$, and all the particles have the mass $m$. Let $\lambda$ and $P$ denote the wavelength and the power of all the trapping lasers, respectively. Furthermore, we assume that the motion of the particles along the plane perpendicular to the optical axis is frozen. We note that the distances between the particles are assumed to be much larger than a characteristic length scale in the collective behavior of particles~\cite{Jones2018,Shahmoon2020,Rusconi2021}.

The interaction between the two particles arises due to the interference between the scattered electromagnetic field and the trapping laser. Since the scattered field acquires the phase $kd_0$ during the propagation, the phase difference between the trapping lasers at the positions of the particles leads to the constructive and destructive interference depending on the propagation direction of the scattered field. It is this spatial asymmetry that renders the interparticle coupling nonreciprocal. We note that the effective open boundaries can be realized by arranging the tightly localized particles at the ends of the system (cf. Fig.~\ref{figappc1} in Appendix~\ref{secC}); the non-Hermitian skin effect then manifests itself as the large oscillation amplitudes close to the boundary regions. Due to the long-range nature of the dipole-dipole interaction, it is in general necessary to incorporate the couplings that reach up to $N$th nearest neighbor particles. Altogether, the linearized equation of motion of the $n$th particle along the $z$ axis is given by
\begin{eqnarray}
&&m\ddot{z}_n+m\gamma\dot{z}_n=-\left(m\Omega^2+2\sum_{l=1}^NK_l\right)z_n \nonumber\\
&&+\sum_{l=1}^N\left[\left(K_l+\bar{K}_l\right)z_{n-l}+\left(K_l-\bar{K}_l\right)z_{n+l}\right].
\label{eq1}
\end{eqnarray}
Here, $\Omega$ is an intrinsic mechanical frequency of the particle proportional to $\sqrt{P}$, $\gamma$ is a friction coefficient, and $K_l$ and $\bar{K_l}$ are the coupling strengths given by
\begin{eqnarray}
\left\{ \begin{array}{l}
\displaystyle K_l=\frac{G}{lk_0d_0}\cos\left(lk_0d_0\right)\cos\left(l\Delta\phi\right), \vspace{3pt}\\
\displaystyle \bar{K}_l=\frac{G}{lk_0d_0}\sin\left(lk_0d_0\right)\sin\left(l\Delta\phi\right),
\end{array}\right.
\label{eq2}
\end{eqnarray}
where $G$ has the dimension of a spring constant and is proportional to $P$, $\Delta\phi$ is the optical phase difference between the nearest neighbor trapping lasers in the focal plane (Fig.~\ref{fig1}), and $k_0\left(=2\pi/\lambda\right)$ is the wavenumber of the trapping laser. We explain the detailed derivation of Eq.~(\ref{eq1}) in Appendix~\ref{secA}. Importantly, the optical phase difference gives rise to the nonreciprocal couplings due to the nonzero $\bar{K}_l$. Thus, our setup is distinct from the array of levitated nanoparticles proposed in Ref.~\cite{Liu2020}, which has investigated a non-Hermitian transport phenomenon with reciprocal couplings. Furthermore, one can infer from Eq.~(\ref{eq2}) that the couplings are long-range because the dipole-dipole interaction is proportional to the inverse of the distance between the particles. We note that the model can be mathematically mapped to a tight-binding model with gain and loss via a similarity transformation~\cite{Okuma2020}.

The continuum bands of non-Hermitian tight-binding models can be obtained by invoking the non-Bloch band theory (cf. Appendix~\ref{secB}), which reproduces the eigenvalues under open boundary conditions. Specifically, the continuum bands are calculated from the generalized Brillouin zone spanned by $\beta\equiv e^{ik}$ for the complex Bloch wavenumber $k$. We here apply the non-Bloch band theory to the levitated nanoparticle array; throughout this paper, we assume $\left|K_N\right|\neq\left|\bar{K}_N\right|$. By substituting $z_n=\psi_ne^{i\omega t}$ to Eq.~(\ref{eq1}), the real-space eigenequation reads
\begin{eqnarray}
&&\frac{1}{m}\sum_{l=1}^N\left[\left(K_l-\bar{K}_l\right)\psi_{n+l}\left(K_l+\bar{K}_l\right)\psi_{n-l}\right] \nonumber\\
&&+\left(\omega^2-i\gamma\omega-\Omega^2-\frac{2}{m}\sum_{l=1}^NK_l\right)\psi_n=0.
\label{eq3}
\end{eqnarray}
Importantly, the ansatz of Eq.~(\ref{eq3}) can be taken as
\begin{equation}
\psi_n=\sum_{j=1}^{2N}\left(\beta_j\right)^n\phi^{\left(j\right)},
\label{eq4}
\end{equation}
where $\beta_j\left(=\beta\right)$ is the solution of the characteristic equation given by
\begin{eqnarray}
&&\frac{1}{m}\sum_{l=1}^N\left[\left(K_l-\bar{K}_l\right)\beta^l+\left(K_l+\bar{K}_l\right)\beta^{-l}\right] \nonumber\\
&&+\left(\omega^2-i\gamma\omega-\Omega^2-\frac{2}{m}\sum_{l=1}^NK_l\right)=0.
\label{eq5}
\end{eqnarray}
We note that Eq.~(\ref{eq5}) is an algebraic equation for $\beta$ of $2N$th degrees. The main result of the non-Bloch band theory is that the condition for the generalized Brillouin zone is obtained from the $2N$ solutions as follows:
\begin{equation}
\left|\beta_N\right|=\left|\beta_{N+1}\right|
\label{eq6}
\end{equation}
with $\left|\beta_1\right|\leq\cdots\leq\left|\beta_{2N}\right|$. The trajectories of $\beta_N$ and $\beta_{N+1}$ form the generalized Brillouin zone on the complex plane, which reveals the essential features of non-Hermitian systems (see, e.g., Refs.~\cite{Yi2020,Okuma2021,Xue2021,Yokomizo2021v2}). Then, we can calculate the continuum bands by combining Eq.~(\ref{eq5}) with the generalized Brillouin zone.

%
% -------------------------------------------------------------------------------------------------------------------------------------------------------------------------------------
%

\section{\label{sec3}Dynamical phase diagram}
\begin{figure}[h!]
\includegraphics[width=7cm]{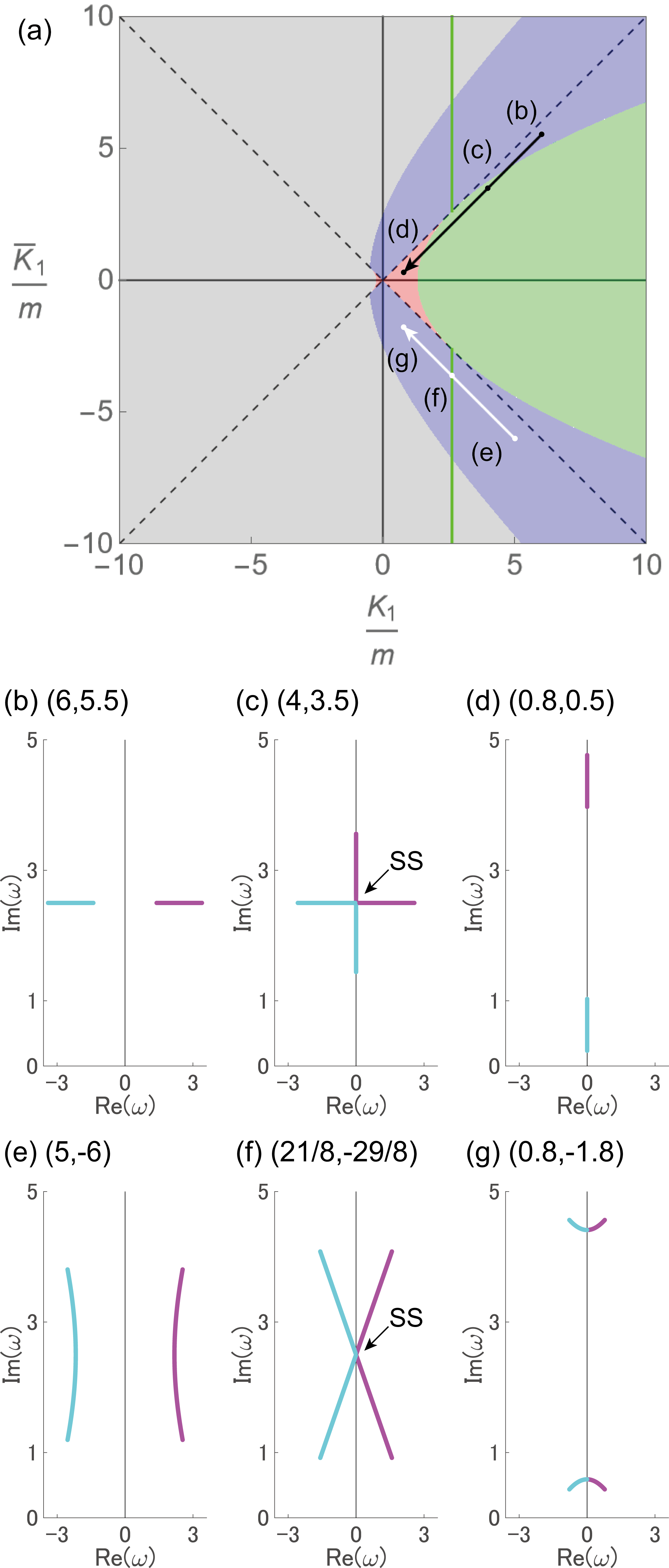}
\caption{\label{fig2} Dynamical phase diagram and continuum bands of the levitated nanoparticle array. (a) Dynamical phase diagram exhibiting the underdamped, critical, overdamped, and dynamically unstable phases shown in the blue, green, red, and gray-shaded regions, respectively. The spectral singularity (SS) appears in the green-shaded region and on the green lines. $\left|K_1\right|=\left|\bar{K}_1\right|$ is satisfied on the black dashed lines. We set the parameters to be $\Omega=1$ and $\gamma=5$. (b--g) Evolutions of the continuum bands along the arrows in panel (a). The magenta and cyan express $\omega_+$ and $\omega_-$, respectively. The numerical values in each panel specify $\left(K_1,\bar{K}_1\right)$.}
\end{figure}
We start our analysis from the levitated nanoparticle array with the nearest-neighbor interaction, which corresponds to $N=1$ in Eq.~(\ref{eq1}); in the following, we assume $\gamma>2\Omega$ for the sake of concreteness. From Eq.~(\ref{eq6}), the generalized Brillouin zone can be given by the circle with the radius $r=\sqrt{\left|\left(K_1+\bar{K}_1\right)/\left(K_1-\bar{K}_1\right)\right|}$. The analytical form of the continuum bands reads
\begin{equation}
\omega_\pm=\frac{i}{2}\gamma\pm\sqrt{\Omega^2+\frac{2}{m}\left(K_1-\sqrt{K_1^2-\bar{K}_1^2}\cos\theta\right)-\frac{\gamma^2}{4}},
\label{eq7}
\end{equation}
where $\theta$ is a real number. Since each eigenmode contributes to the dynamics through the factor $e^{-{\rm Im}\left(\omega_\pm\right)}e^{i{\rm Re}\left(\omega_\pm\right)}$, we can show the dynamical phase diagram depending on $K_1/m$ and $\bar{K}_1/m$ [Fig.~\ref{fig2}(a)]. We here emphasize that the continuum bands discussed here can have direct experimental relevance. Indeed, the theoretical calculation of the eigenvalues in the coupled two levitated nanoparticles has explained well experimentally observed crossing/avoided crossing of the eigenspactra and the appearance of an exceptional point~\cite{Rieser2022}. Figures~\ref{fig2}(b)--\ref{fig2}(d) and \ref{fig2}(e)--\ref{fig2}(g) plot the evolutions of the continuum bands along the black and white arrows indicated in Fig.~\ref{fig2}(a), respectively.

In the blue-shaded regions of Fig.~\ref{fig2}(a), all the particles oscillate with the attenuation because ${\rm Re}\left(\omega_\pm\right)\neq0$ and ${\rm Im}\left(\omega_\pm\right)>0$ [Fig.~\ref{fig2}(b)]. In contrast, in the red-shaded regions, their motion monotonically  vanishes without oscillations because ${\rm Re}\left(\omega_\pm\right)=0$ and ${\rm Im}\left(\omega_\pm\right)>0$ [Fig.~\ref{fig2}(d)]. For these reasons, we term the former (latter) the dynamical phase as the underdamped (overdamped) phase.

Remarkably, we find the broad green-shaded region where the two branches coalesce at ${\rm Re}\left(\omega_\pm\right)=0$ [Fig.~\ref{fig2}(c)]; this degeneracy is called the spectral singularity. There, we find the crossover behavior where the overdamped behavior eventually sets in after the initial underdamped oscillations; we shall term this intermediate regime as the critical phase. One of its key characteristics is the presence of the particles near the boundaries which are strongly driven by the adjacent trapping lasers inducing the nonreciprocal couplings. It is worthwhile to mention that, with $\gamma<2\Omega$, the critical phase appears on the parameter region where the sign of $K/m$ becomes negative [cf. Fig.~\ref{figappc2}(a) in Appendix~\ref{secC}]. The transient phenomenon discussed here is supported by the non-Hermitian skin effect, since the critical phase disappears under periodic boundary conditions, as discussed in Appendix~\ref{secD}. As shown in Fig.~\ref{fig2}(f), the spectral singularity also appears along the green vertical lines in Fig.~\ref{fig2}(a). However, one would need the fine-tuning of the parameters in this case as indicated by Figs.~\ref{fig2}(e)--\ref{fig2}(g), where the two continuum bands are recombined across the green line.

Additionally, we also find the dynamically unstable phase as indicated by the gray-shaded region in Fig.~\ref{fig2}(a). There, the driving forces from the adjacent trapping lasers give rise to the dynamical instability of the particles where the oscillation amplitudes diverge in the long-time limit, because negative hopping amplitudes cause the force that increasingly keeps away the particles from their equilibrium positions. We expect that nonlinear effects will eventually play a crucial role in this phase, since the amplification is eventually balanced by nonlinear suppression. The dynamically unstable phase discussed here is difficult to realize in previous non-Hermitian systems due to the lack of the ability to implement negative hopping amplitudes. It is worthwhile to mention that, in finite-size systems, the phase boundary between the overdamped and dynamically unstable phases can be slightly modified, as discussed in Appendix~\ref{secC}.

%
% -------------------------------------------------------------------------------------------------------------------------------------------------------------------------------------
%

\section{\label{sec4}Nonreciprocal long-range interaction}
\begin{figure}[h!]
\includegraphics[width=8.5cm]{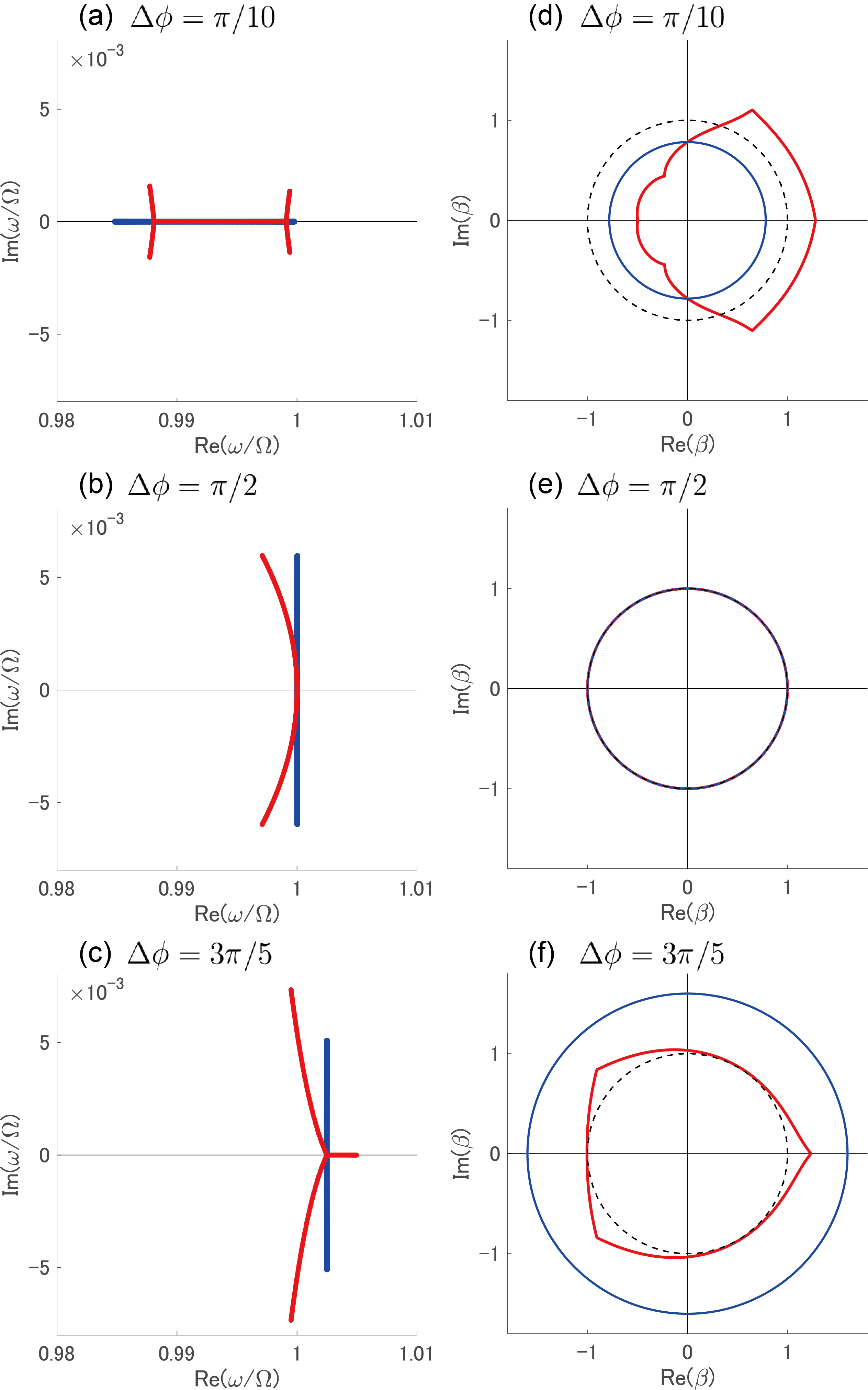}
\caption{\label{fig3}Continuum bands and generalized Brillouin zones of the levitated nanoparticle array at different $\Delta\phi$. (a--c) The continuum bands with the positive branch of the square root, and (d--f) the corresponding generalized Brillouin zones are shown. The red (blue) curves indicate the results for $N=2$ ($N=1$). In panels (d--f), the black dashed curve expresses the conventional Brillouin zone spanned by $\beta\equiv e^{ik}~\left(k\in{\mathbb R}\right)$. The system parameters are set to be $\lambda=1.064\times10^{-6}~{\rm m},d_0=10^{-5}~{\rm m},\Omega=10^5~{\rm s^{-1}}$, and $G/\left(mk_0d_0\right)=10^8~{\rm s^{-2}}$.}
\end{figure}
We next investigate how the long-range nature of the couplings can affect the continuum band and the corresponding generalized Brillouin zone; in the following, we neglect the friction for the sake of simplicity. We assume that the interaction reaches up to the next-nearest-neighbor particles, which corresponds to $N=2$ in Eq.~(\ref{eq1}). In Fig.~\ref{fig3}, we plot the continuum bands with the positive branch of the square root and the corresponding generalized Brillouin zone at different $\Delta\phi$. The black dashed curves in Figs.~\ref{fig3}(d)--(f) indicate the conventional Brillouin zone formed by $\beta\equiv e^{ik}~\left(k\in\mathbb R\right)$. We note that the parameters considered in these calculations satisfy $K_1/m,\bar{K}_1/m,K_2/m,\bar{K}_2/m\ll\Omega^2$ and have been experimentally realized in Ref.~\cite{Rieser2022} in the case of two particles.

One can see from Figs.~\ref{fig3}(d) and \ref{fig3}(f) that the generalized Brillouin zone with $N=2$ forms a skewed closed curve with the cusps, at which it becomes indifferentiable, while the generalized Brillouin zone with $N=1$ is merely a circle. Importantly, the cusps correspond to the self-crossing points of the continuum band [Figs.~\ref{fig3}(a) and \ref{fig3}(c)]~\cite{Yokomizo2020v2}. Thus, the long-range nature of the nonreciprocal couplings can lead to these unconventional band structures. Meanwhile, at $\Delta\phi=\pi/2$, the generalized Brillouin zone becomes the unit circle independently of $N$ [Fig.~\ref{fig3}(e)], and there are no self-crossing points [Fig.~\ref{fig3}(b)], where the non-Hermitian skin effect disappears.

It is noteworthy that, in the case of Fig.~\ref{fig2}(c), nonorthogonality of the eigenstates with the eigenvalues around the self-crossing points is stronger than that of the other eigenstates away from the self-crossing point. This is because the overlap of the left and right eigenvectors becomes minimum at the self-crossing point. Thereby, such eigenstates exhibit a striking response against perturbations~\cite{Schomerus2020}, which indicates that the excitation modes around the self-crossing point could be utilized for a highly sensitive sensor.

%
% -------------------------------------------------------------------------------------------------------------------------------------------------------------------------------------
%

\section{\label{sec5}Summary and Discussion}
In summary, we propose and analyze the levitated nanoparticle array as an ideal platform to study new realms of non-Hermitian physics in a highly controlled manner. We show that the system exhibits the unconventional critical phase, where the spectral singularity originating from the non-Hermitian skin effect persists over a broad parameter region. We also point out that the tunable dipole-dipole interaction allows for extremely nonreciprocal hopping amplitudes with possibly negative signs, which result in the dynamical instability. We finally reveal that the long-range nature of the couplings further enriches the non-Hermitian band structures, leading to the cusps of the generalized Brillouin zone and the self-crossing point of the continuum band.

We expect that the levitated nanoparticle array will be an ideal platform to explore rich phenomena induced by nonreciprocal long-range hopping amplitudes thanks to its high tunability. For example, a nontrivial topological phase transition mediated by a topological semimetal phase with exceptional points has been proposed~\cite{Yokomizo2020v2,Xu2021}. Additionally, the number of branches from the self-crossing point of the continuum band increases as a coupling distance between two particles~\cite{Zeng2022}. Furthermore, it has been proposed that nonreciprocal long-range couplings give rise to bulk eigenstates which exhibit the crossover from a constant localization length to a system-size-dependent localization length~\cite{Wang2023}. This previous work has also indicated that, since long-range couplings suppress the non-Hermitian skin effect, the scaling of entanglement entropy can change from an area law to a subextensive law.

The continuum bands of the levitated nanoparticle array can be experimentally studied by measuring the power spectral density. It is thus feasible to directly observe the spectral singularity and the self-crossing points. Meanwhile, to access the strong coupling regime with $K/m\Omega^2={\rm O}\left(1\mathchar`-10\right)$ considered in Fig.~\ref{fig2}, it is necessary to realize a stronger dipole-dipole interaction than the one already realized in Ref.~\cite{Rieser2022}. We expect that this should be made possible by increasing size of the particles and using lasers with a shorter wavelength. Also, we note that the friction coefficient can be controlled by changing the pressure of the surrounding gas, while thermal noise can be suppressed by keeping its temperature sufficiently low. We discuss in detail the experimental feasibility in Appendix~\ref{secE}.

It is interesting to further explore various aspects and potentials of a levitated nanoparticle array, since this unique platform opens a new avenue of investigating sensing, nonlinearity, and nonequilibrium quantum physics. First, the critical phase can be potentially applicable for enhanced sensing because the strong nonorthogonality associated with the spectral singularity which one can observe in the array of a few particles (cf. see Appendix~\ref{secC}) is known to trigger the singular sensitivity to perturbation~\cite{Hodaei2017,Chen2017,Hokmabadi2019}. Second, it is crucial to reveal the competition between nonreciprocity and nonlinearity~\cite{Fruchart2021}, where a levitated nanoparticle array is expected to exhibit rich collective phenomena such as synchronization. Third, an extension of the present analysis to quantum regimes is an intriguing open problem. For instance, it merits further study to understand how the nonreciprocal couplings affect the entanglement between particles~\cite{Rudolph2022}, how the dynamical instability of the levitated nanoparticle array can be utilized to squeeze the mechanical mode of a levitated nanoparticle~\cite{Kustura2022}, and whether or not the nonorthogonality is helpful for quantum metrology~\cite{Weiss2021}.

%
% -------------------------------------------------------------------------------------------------------------------------------------------------------------------------------------
%

\begin{acknowledgments}
K.Y. was supported by JSPS KAKENHI through Grant No.~JP21J01409. Y.A. acknowledges support from the Japan Society for the Promotion of Science through Grant No.~JP19K23424 and from JST FOREST Program (Grant No.~JPMJFR222U, Japan).
\end{acknowledgments}

%
% -------------------------------------------------------------------------------------------------------------------------------------------------------------------------------------
%

\appendix

%
% -------------------------------------------------------------------------------------------------------------------------------------------------------------------------------------
%

\section{\label{secA}Equation of motion}
\begin{figure}[]
\includegraphics[width=8.5cm]{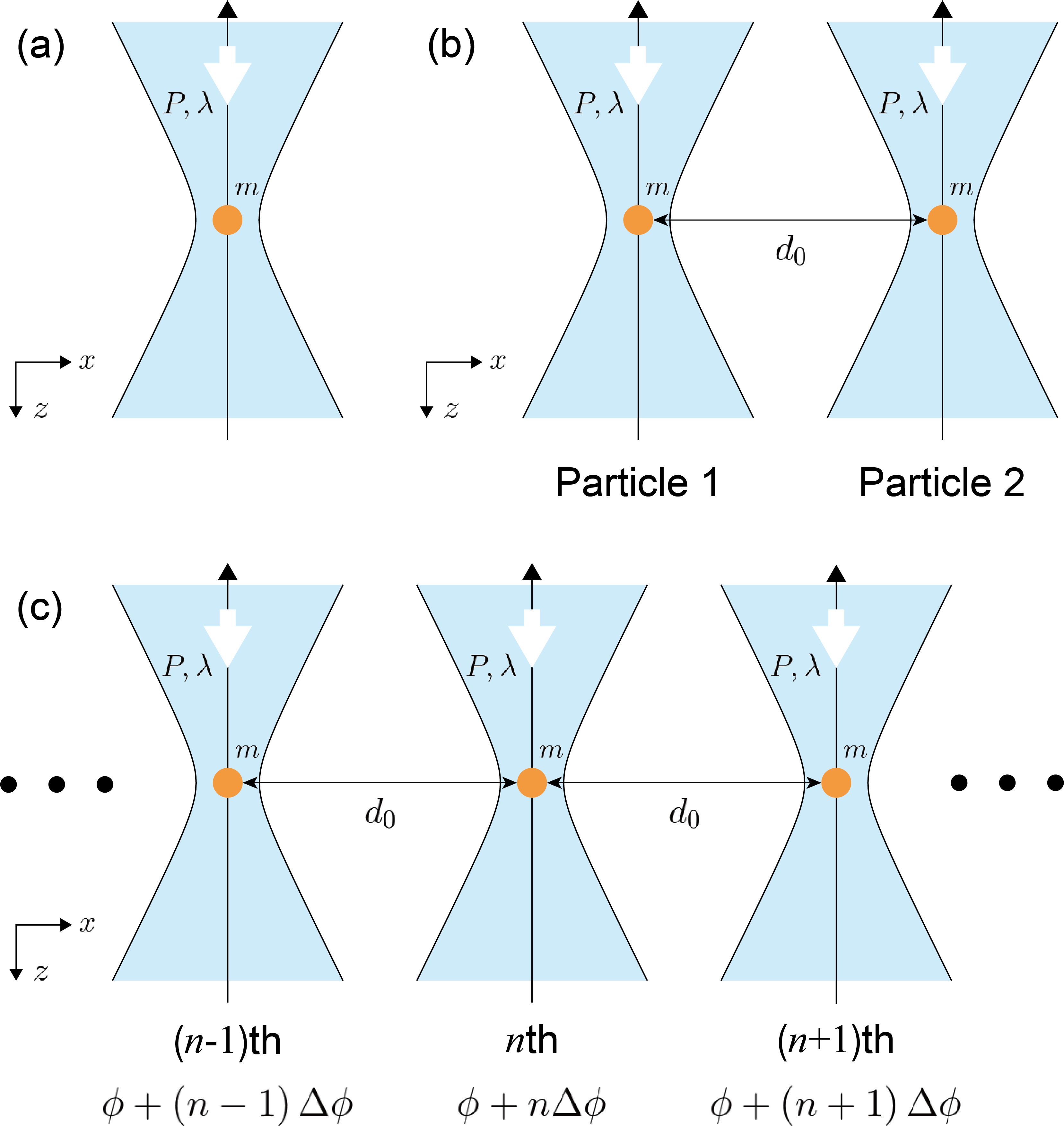}
\caption{\label{figappa}Schematic figure of a single levitated nanoparticle, coupled two levitated nanoparticles, and an array of leviated nanoparticles. In panels (a), (b), and (c), the mass of the particle is $m$, and the trapping lasers have the power $P$ and the wavelength $\lambda$. In panels (b) and (c), the distance between the particles is $d_0$. In panel (c), we set the phase of the $n$th trapping laser in the focal plane to be $\phi+n\Delta\phi$.}
\end{figure}
We derive the equation of motion of a single levitated nanoparticle, coupled two levitated nanoparticles, and an array of levitated nanoparticles as shown in Figs.~\ref{figappa}(a), \ref{figappa}(b), and \ref{figappa}(c), respectively. We assume that the motion of the particles along the plane perpendicular to the optical axis is frozen.

%
% -------------------------------------------------------------------------------------------------------------------------------------------------------------------------------------
%

\subsection{\label{secA-1}Single levitated nanoparticle}
We first focus on the motion of a single levitated nanoparticle along the $z$ axis [Fig.~\ref{figappa}(a)]. Let $P$ and $\lambda$ denote the power and the wavelength of the trapping laser, respectively. In this system, the subwavelength-sized particle with the refractive index $n$ is surrounded by a medium with the refractive index $n^\prime$, and it can be regarded as a point dipole with the electric dipole moment given by
\begin{equation}
{\bm p}\left({\bm r},t\right)=\alpha{\bm E}\left({\bm r},t\right).
\label{eqappa1}
\end{equation}
Here, ${\bm E}\left({\bm r},t\right)$ expresses the electric field of the trapping laser, and $\alpha$ is the polarizability of the particle. The latter can explicitly be written as
\begin{equation}
\alpha=3\varepsilon_0n^{\prime2}V\frac{n_r^2-1}{n_r^2+2},
\label{eqappa2}
\end{equation}
where $c$ is the speed of light in vacuum, $V$ is the volume of the particle, and $n_r\equiv n/n^\prime$ is the relative refractive index of the particle. The electric field then acts on the point dipole, and the particle feels the gradient force~\cite{Harada1996} given by
\begin{eqnarray}
{\bm F}_{\rm grad}\left({\bm r}\right)&=&\left<\frac{1}{2}{\rm Re}\left[\left(\alpha{\bm E}^\ast\left({\bm r},t\right)\cdot{\bm\nabla}\right){\bm E}\left({\bm r},t\right)\right]\right>_T \nonumber\\
&=&\frac{\alpha}{2c\varepsilon_0n^\prime}{\bm\nabla}I\left({\bm r}\right),
\label{eqappa3}
\end{eqnarray}
where $\left<\cdots\right>_T$ means time average, and $I\left({\bm r}\right)$ is the spatial profile of the intensity of the trapping laser. We note that the trapping laser is described by the Gaussian beam, of which the electric field is given by
\begin{eqnarray}
{\bm E}\left({\bm r},t\right)&=&{\bm E}_0\frac{w_0}{w\left(z\right)} \nonumber\\
&&\times\exp\left[-\frac{x^2+y^2}{\left\{w\left(z\right)\right\}^2}+ik\frac{x^2+y^2}{2R\left(z\right)}-i\zeta\left(z\right)\right] \nonumber\\
&&\times e^{i\left(k_0z-\omega t\right)},
\label{eqappa4}
\end{eqnarray}
where
\begin{eqnarray}
\left\{ \begin{array}{l}
\displaystyle w\left(z\right)=w_0\sqrt{1+\left(\frac{z}{z_0}\right)^2}, \vspace{5pt}\\
\displaystyle R\left(z\right)=z\left[1+\left(\frac{z_0}{z}\right)^2\right], \vspace{5pt}\\
\displaystyle \zeta\left(z\right)=\tan^{-1}\left(\frac{z}{z_0}\right).
\end{array}\right.
\label{eqappa5}
\end{eqnarray}
Here, $k_0\left(=2\pi/\lambda\right)$ is the wavenumber of the trapping laser, $w_0$ is the beam waist, and $z_0=\pi w_0^2/\lambda$ is the Rayleigh length. The combination of Eqs.~(\ref{eqappa3}) and (\ref{eqappa4}) leads to the $z$ component of the gradient force given by
\begin{equation}
F_{{\rm grad},z}\left(z\right)=-\frac{P\alpha}{2\pi c\varepsilon_0n^\prime w_0^2z_0^2}\frac{z}{\left[1+\left(z/z_0\right)^2\right]^2}.
\label{eqappa6}
\end{equation}
In the vicinity of the focal plane of the trapping laser, the linearized equation of the motion of the levitated nanoparticle is obtained as
\begin{equation}
m\ddot{z}+m\gamma\dot{z}=-m\Omega^2z,
\label{eqappa7}
\end{equation}
where
\begin{equation}
\Omega^2=\frac{P\alpha}{2\pi c\varepsilon_0n^\prime mw_0^2z_0^2},
\label{eqappa8}
\end{equation}
and $\gamma$ is a friction coefficient. We remark that the linearized motion of the particle exhibits the harmonic oscillation because the gradient force plays a role as the restoring force.

%
% -------------------------------------------------------------------------------------------------------------------------------------------------------------------------------------
%

\subsection{\label{secA-2}Coupled two levitated nanoparticles}
We next consider the equation of motion of the coupled two levitated nanoparticles [Fig.~\ref{figappa}(b)]. We assume that the polarization direction of the particles points to the $y$ axis. Then, in addition to the gradient force, both particles feel the interaction caused by the dipole radiation, which is called the optical binding force~\cite{Dapasse1994,Dholakia2010}.

First of all, we shall explain the mechanism of the optical binding force based on Ref.~\cite{Rieser2022}. Importantly, there exist two contributions to the optical binding force. The first contribution results from the combination of the electric field scattered from one particle to another particle and the electric dipole induced by the trapping laser. The second contribution originates from the acting of the trapping laser on the electric dipole induced by the scattered field. Specifically, for particle 1, the sum of these contributions leads to the form of the optical binding force as follows:
\begin{eqnarray}
&&{\bm F}_{\rm bind}^{2\rightarrow1}\left({\bm r}_1,{\bm r}_2\right) \nonumber\\
&&=\left<\frac{1}{2}{\rm Re}\left[\left(\alpha{\bm E}^\ast\left({\bm r}_1,t\right)\cdot{\bm\nabla}_{{\bm r}_1}\right){\bm E}_{\rm sca}^{2\rightarrow1}\left({\bm r}_1,{\bm r}_2,t\right)\right]\right. \nonumber\\
&&\left.+\frac{1}{2}{\rm Re}\left[\left(\alpha{\bm E}_{\rm sca}^{2\rightarrow1\ast}\left({\bm r}_1,{\bm r}_2,t\right)\cdot{\bm\nabla}_{{\bm r}_1}\right){\bm E}\left({\bm r}_1,t\right)\right]\right>_T. \nonumber\\
\label{eqappa9}
\end{eqnarray}
Here, ${\bm E}_{\rm sca}^{2\rightarrow1}\left({\bm r}_1,{\bm r}_2,t\right)$ is the electric field scattered from particle 2 to particle 1, and its form is written as
\begin{equation}
{\bm E}_{\rm sca}^{2\rightarrow1}\left({\bm r}_1,{\bm r}_2,t\right)=G\left({\bm r}_1-{\bm r}_2\right)\alpha{\bm E}\left({\bm r}_2,t\right),
\label{eqappa10}
\end{equation}
where $G\left({\bm r}\right)$ called the Green's tensor is the electric field propagator between the two dipoles~\cite{Jackson1999}, and it is given by
\begin{equation}
G\left({\bm r}\right)=\frac{e^{ik_0r}}{4\pi\varepsilon_0}\left[\frac{3{\bm r}\otimes{\bm r}-r^2}{r^5}\left(1-ik_0r\right)+k_0^2\frac{r^2-{\bm r}\otimes{\bm r}}{r^3}\right].
\label{eqappa11}
\end{equation}
By using
\begin{eqnarray}
&&{\rm Re}\left[\left({\bm\nabla}_{{\bm r}_1}\times{\bm E}\left({\bm r}_1,t\right)\right)\times{\bm E}_{\rm sca}^{2\rightarrow1\ast}\left({\bm r}_1,{\bm r}_2,t\right)\right] \nonumber\\
&&-{\rm Re}\left[{\bm E}_{\rm sca}^{2\rightarrow1}\left({\bm r}_1,{\bm r}_2,t\right)\times\left({\bm\nabla}_{{\bm r}_1}\times{\bm E}^\ast\left({\bm r}_1,{\bm r}_2,t\right)\right)\right] \nonumber\\
&&={\rm Re}\left[\left({\bm E}^\ast\left({\bm r}_1,t\right)\cdot{\bm\nabla}_{{\bm r}_1}\right){\bm E}_{\rm sca}^{2\rightarrow1}\left({\bm r}_1,{\bm r}_2,t\right)\right] \nonumber\\
&&+{\rm Re}\left[\left({\bm E}_{\rm sca}^{2\rightarrow1\ast}\left({\bm r}_1,{\bm r}_2,t\right)\cdot{\bm\nabla}_{{\bm r}_1}\right){\bm E}\left({\bm r}_1,t\right)\right] \nonumber\\
&&-{\rm Re}\left[{\bm\nabla}_{{\bm r}_1}\left({\bm E}^\ast\left({\bm r}_1,t\right)\cdot{\bm E}_{\rm sca}^{2\rightarrow1}\left({\bm r}_1,{\bm r}_2,t\right)\right)\right]
\label{eqappa12}
\end{eqnarray}
and ${\bm\nabla}_{{\bm r}_1}\times{\bm E}\left({\bm r}_1,t\right)={\bm0}$, we can rewrite Eq.~(\ref{eqappa9}) to a brief form given by
\begin{eqnarray}
&&{\bm F}_{\rm bind}^{2\rightarrow1}\left({\bm r}_1,{\bm r}_2\right) \nonumber\\
&&=\left<\frac{1}{2}{\bm\nabla}_{{\bm r}_1}{\rm Re}\left[\alpha{\bm E}^\ast\left({\bm r}_1,t\right)G\left({\bm r}_1-{\bm r}_2\right)\alpha{\bm E}\left({\bm r}_2,t\right)\right]\right>_T. \nonumber\\
\label{eqappa13}
\end{eqnarray}
In the following, we calculate the $z$ component of the optical binding force. In the far-field regime with $k_0d_0\gg1$, the dominant contribution from the Green's tensor to the optical binding force is the term proportional to $1/r$. Furthermore, in the vicinity of the focal plane, the $z$ dependence of the electric field is approximated as
\begin{equation}
{\bm E}\left({\bm r}_j,t\right)\approx{\bm E}_0\exp\left[i\phi_j+i\left(k_0-\frac{1}{z_0}\right)z_j-i\omega t\right]
\label{eqappa14}
\end{equation}
for $j=1,2$, where $\phi_j$ expresses the optical phase at the focal plane. Substituting Eqs.~(\ref{eqappa11}) and (\ref{eqappa14}) into Eq.~(\ref{eqappa13}), we can get the optical binding force along the $z$ axis as follows:
\begin{eqnarray}
&&F_{{\rm bind},z}^{2\rightarrow1}\left(z_1,z_2\right)\approx\frac{P\alpha^2k_0^3\left(k_0-1/z_0\right)}{2\pi^2c\varepsilon_0^2n^\prime w_0^2} \nonumber\\
&&\times\sin\left[k_0d_0-\Delta\phi-\left(k_0-\frac{1}{z_0}\right)\left(z_1-z_2\right)\right], \nonumber\\
\label{eqappa15}
\end{eqnarray}
where $\Delta\phi\equiv\phi_1-\phi_2$.

We can similarly obtain the optical binding force for particle 2 as follows:
\begin{eqnarray}
&&{\bm F}_{\rm bind}^{1\rightarrow2}\left({\bm r}_1,{\bm r}_2\right) \nonumber\\
&&=\left<\frac{1}{2}{\bm\nabla}_{{\bm r}_2}{\rm Re}\left[\alpha{\bm E}^\ast\left({\bm r}_2,t\right)G\left({\bm r}_2-{\bm r}_1\right)\alpha{\bm E}\left({\bm r}_1,t\right)\right]\right>_T. \nonumber\\
\label{eqappa16}
\end{eqnarray}
Therefore, the explicit form of the optical binding force along the $z$ axis is derived as
\begin{eqnarray}
&&F_{{\rm bind},z}^{1\rightarrow2}\left(z_1,z_2\right)\approx\frac{P\alpha^2k_0^3\left(k_0-1/z_0\right)}{2\pi^2c\varepsilon_0^2n^\prime w_0^2} \nonumber\\
&&\times\sin\left[k_0d_0+\Delta\phi+\left(k_0-\frac{1}{z_0}\right)\left(z_1-z_2\right)\right]. \nonumber\\
\label{eqappa17}
\end{eqnarray}

Remarkably, the interaction between these two particles becomes nonreciprocal because $F_{{\rm bind},z}^{2\rightarrow1}\left(z_1,z_2\right)\neq-F_{{\rm bind},z}^{1\rightarrow2}\left(z_2,z_1\right)$. The key ingredient of this nonreciprocity is the interference between the trapping laser and the scattered field. Let $\Phi_j~\left(j=1,2\right)$ denote the optical phase of the trapping laser at the position of the particle. The interference depends on the local phase difference $\Delta\Phi\equiv\Phi_1-\Phi_2$ and the phase accumulation $kd_0$ which the scattered field acquires during the propagation. Specifically, while the contribution of the interference is $kd_0-\Delta\Phi$ within the propagation of the scattered field from particle 1 to particle 2, it becomes $kd_0+\Delta\Phi$ in the opposite case. As a result, the interaction originating from the interference becomes spatially asymmetric.

We can now obtain the linearized equation of motion of the coupled levitated nanoparticles as follows:
\begin{eqnarray}
\left\{ \begin{array}{l}
m\ddot{z}_1+m\gamma\dot{z}_1=-\left(m\Omega^2+K+\bar{K}\right)z_1+\left(K+\bar{K}\right)z_2, \vspace{3pt}\\
m\ddot{z}_2+m\gamma\dot{z}_2=-\left(m\Omega^2+K-\bar{K}\right)z_2+\left(K-\bar{K}\right)z_1,
\end{array}\right. \nonumber\\
\label{eqappa18}
\end{eqnarray}
where $\Omega$ is the intrinsic mechanical frequency given by Eq.~(\ref{eqappa1}). The coupling constants are given by
\begin{eqnarray}
\left\{ \begin{array}{l}
\displaystyle K=\frac{G}{k_0d_0}\cos\left(k_0d_0\right)\cos\left(\Delta\phi\right), \vspace{3pt}\\
\displaystyle \bar{K}=\frac{G}{k_0d_0}\sin\left(k_0d_0\right)\sin\left(\Delta\phi\right),
\end{array}\right.
\label{eqappa19}
\end{eqnarray}
and
\begin{equation}
G=\frac{P\alpha^2k_0^3\left(k_0-1/z_0\right)^2}{2\pi^2c\varepsilon_0^2n^\prime w_0^2}.
\label{eqappa20}
\end{equation}

%
% -------------------------------------------------------------------------------------------------------------------------------------------------------------------------------------
%

\subsection{\label{secA-3}Levitated nanoparticle array}
We finally study the arrangement of the multiple levitated nanoparticles at equal interval $d_0$ [Fig.~\ref{figappa}(c)]. In this system, the dipole-dipole interaction among the several particles arises from the multiple scattering of the trapping lasers. Nevertheless, the dominant contribution to the dynamics of the system comes from the interaction between two particles. Thus, we neglect higher-order scattering processes. This corresponds to approximating the optical binding force up to ${\cal O}\left(\left|{\bm p}\right|^2\right)$.

We shall explain how one can derive the equation of motion of the levitated nanoparticle array. We set the optical phase of the $n$th trapping laser in the focal plane to be $\phi+n\Delta\phi$. For the $n$th and $n+l$th particles, the phase difference between the trapping lasers is $l\Delta\phi$, and the distance between the particles is $ld_0$. Hence, the interaction between these particles can be obtained by the same procedure as explained above. Due to the long-range nature of the dipole-dipole interaction, it is necessary to incorporate the couplings that reach up to $N$th neighbor particles. Then, the equation of motion of the system is written as
\begin{eqnarray}
m\ddot{z}_n+m\gamma\dot{z}_n&=&-\left(m\Omega^2+2\sum_{l=1}^NK_l\right)z_n \nonumber\\
&&+\sum_{l=1}^N\left[\left(K_l+\bar{K}_l\right)z_{n-l}+\left(K_l-\bar{K}_l\right)z_{n+l}\right], \nonumber\\
\label{eqappa21}
\end{eqnarray}
and
\begin{eqnarray}
\left\{ \begin{array}{l}
\displaystyle K_l=\frac{G}{lk_0d_0}\cos\left(lk_0d_0\right)\cos\left(l\Delta\phi\right), \vspace{3pt}\\
\displaystyle \bar{K}_l=\frac{G}{lk_0d_0}\sin\left(lk_0d_0\right)\sin\left(l\Delta\phi\right),
\end{array}\right.
\label{eqappa22}
\end{eqnarray}
where the constant $G$ is given in Eq.~(\ref{eqappa20}).

%
% -------------------------------------------------------------------------------------------------------------------------------------------------------------------------------------
%

\section{\label{secB}Non-Bloch band theory}
We describe the physical meaning of the non-Bloch band theory for 1D non-Hermitian tight-binding models. First of all, we show that the condition for the generalized Brillouin zone can be interpreted as the condition that the ``plane waves'' form the standing wave, by using the specific model. We next prove that the recombination of the ``plane waves'' forming the standing wave occurs at the cusps of the generalized Brillouin zone.

%
% -------------------------------------------------------------------------------------------------------------------------------------------------------------------------------------
%

\subsection{\label{secB-1}Condition for the generalized Brillouin zone}
\begin{figure}[]
\includegraphics[width=8.5cm]{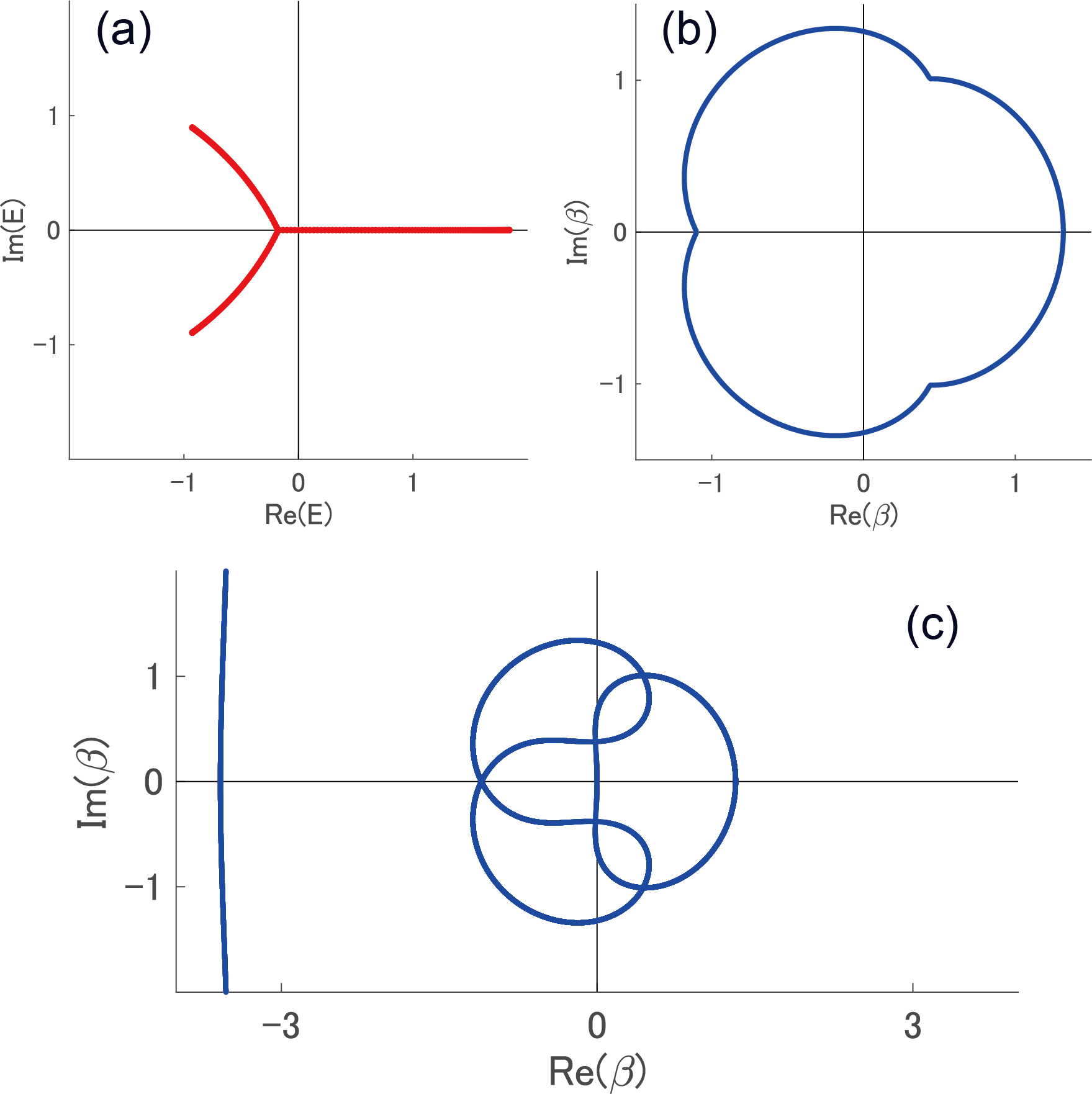}
\caption{\label{figappb}Continuum band and trajectories of the solutions of the characteristic equation of the model (\ref{eqappb1}). (a) Continuum band. (b) Generalized Brillouin zone (c) Sets of the solutions of the characteristic equation satisfying Eq.~(\ref{eqappb7}). We set the parameters to be $t_{+,2}=0.9,t_{+,1}=0.3,t_{-,1}=0.7$, and $t_{-,2}=0.1$.}
\end{figure}

We consider the 1D non-Hermitian tight-binding model with asymmetric hopping amplitudes, the Hamiltonian of which reads
\begin{eqnarray}
H&=&\sum_n\left(t_{+,2}c_{n+2}^\dag c_n+t_{+,1}c_{n+1}^\dag c_n\right. \nonumber\\
&&\left.+t_{-,1}c_n^\dag c_{n+1}+t_{-,2}c_n^\dag c_{n+2}\right),
\label{eqappb1}
\end{eqnarray}
where all the parameter are real numbers. The Schr\"{o}dinger equation can be then written in the form of the real-space eigenequation given by
\begin{equation}
t_{+,2}\psi_{n-2}+t_{+,1}\psi_{n-1}+t_{-,1}\psi_{n+1}+t_{-,2}\psi_{n+2}=E\psi_n,
\label{eqappb2}
\end{equation}
where $\psi_n$ means the amplitude of the state at the site $n$. A general difference theory allows us to take the form of the linear combination
\begin{equation}
\psi_n=\sum_j\left(\beta_j\right)^n\phi^{\left(j\right)},
\label{eqappb3}
\end{equation}
which corresponds to the plane-wave expansion, as the ansatz of Eq.~(\ref{eqappa2}), due to the spatial periodicity. Here, $\beta_j\left(=\beta\right)$ is the solution of the characteristic equation written as
\begin{equation}
t_{+,2}\beta^{-2}+t_{+,1}\beta^{-1}+t_{-,1}\beta+t_{-,2}\beta^2=E,
\label{eqappb4}
\end{equation}
which is a quadratic equation for $\beta$. We note that it is necessary to combine the open boundary conditions obtained from Eqs.~(\ref{eqappb3}) and Eq.~(\ref{eqappb4}) to get the asymptotic set of the energy eigenvalues in the thermodynamic limit. Although the calculation is cumbersome for a large system size, the non-Bloch band theory allows us to avoid the procedure of calculating the continuum band~\cite{Yokomizo2019,Yokomizo2020}. It has been shown that the values of $\beta$ are restricted to lie on the closed curve so that the wavefunction satisfies the open boundary conditions. The closed curve, called the generalized Brillouin zone, is then formed by $\beta=e^{ik}$ for the Bloch wavenumber $k$. It is important that, for the four solutions of Eq.~(\ref{eqappb4}), the condition for the generalized Brillouin zone reads
\begin{equation}
\left|\beta_2\right|=\left|\beta_3\right|
\label{eqappb5}
\end{equation}
with
\begin{equation}
\left|\beta_1\right|\leq\left|\beta_2\right|\leq\left|\beta_3\right|\leq\left|\beta_4\right|.
\label{eqappb6}
\end{equation}
We note that the trajectories of $\beta_2$ and $\beta_3$ satisfying Eq.~(\ref{eqappb5}) form the generalized Brillouin zone. Finally, the combination of Eqs.~(\ref{eqappb4}) and (\ref{eqappb5}) gives the continuum band of the system. We exemplify the continuum band and the generalized Brillouin zone with the specific parameters as shown in Figs.~\ref{figappb}(a) and \ref{figappb}(b).

The Bloch wavenumber obtained by Eq.~(\ref{eqappb5}) becomes complex numbers, which indicates that the bulk eigenstates are localized at boundaries of the system due to the non-Hermitian skin effect. Meanwhile, the generalized Brillouin zone forms a unit circle, when the system becomes a Hermitian system: $t_{+,1}=t_{-,1}$ and $t_{+,2}=t_{-,2}$. This means that the Bloch wavenumber takes real values in Hermitian tight-binding systems, which is consistent with the result of the conventional Bloch band theory.

It is remarkable that the condition for the generalized Brillouin zone can be interpreted in the viewpoint of the non-Hermitian skin effect. The physical meaning of Eq.~(\ref{eqappb5}) is that the localization lengths of the ``plane waves'' corresponding to $\beta_2$ and $\beta_3$ match each other, which leads to the formation of the standing wave by the interference of the ``plane waves''. Furthermore, it is intriguing that the asymptotic energy eigenvalues of the open chain in the thermodynamic limit do not depend on any boundary conditions, since Eqs.~(\ref{eqappb4}) and (\ref{eqappb5}) are independent of boundary conditions of the open chain.

%
% -------------------------------------------------------------------------------------------------------------------------------------------------------------------------------------
%

\subsection{\label{secB-2}Cusps of the generalized Brillouin zone}
We nest explain the appearance mechanism of the cusp of the generalized Brillouin zone, at which it becomes indifferentiable. To this end, we investigate what happens if we impose
\begin{equation}
\left|\beta_i\right|=\left|\beta_j\right|
\label{eqappb7}
\end{equation}
to the system, for some $i$ and $j$ among the four solutions of Eq.~(\ref{eqappb4}). We then show several sets of $\beta_i$ and $\beta_j$ satisfying Eq.~(\ref{eqappb7}) in Fig.~\ref{figappb}(c). We note that the trajectory satisfying Eq.~(\ref{eqappb5}) among the sets is equivalent to the generalized Brillouin zone. Compared with Figs.~\ref{figappb}(b) and (c), one can see that the cusps appear, when the three of the four solutions share the same absolute values. Namely, for $\left|\beta_1\right|<\left|\beta_2\right|=\left|\beta_3\right|<\left|\beta_4\right|$, $\left|\beta_1\right|$ approaches $\left|\beta_2\right|\left(=\left|\beta_3\right|\right)$ as we go around the generalized Brillouin zone, and the behavior of the solutions satisfying Eq.~(\ref{eqappb5}) eventually changes when $\left|\beta_1\right|=\left|\beta_2\right|=\left|\beta_3\right|$. Thus, the recombination of the ``plane waves'' forming the standing wave occurs at the cusps of the generalized Brillouin zone and the corresponding self-crossing points of the continuum band.

%
% -------------------------------------------------------------------------------------------------------------------------------------------------------------------------------------
%

\section{\label{secC}Array of finite levitated nanoparticles}
\begin{figure*}[]
\includegraphics[width=16cm]{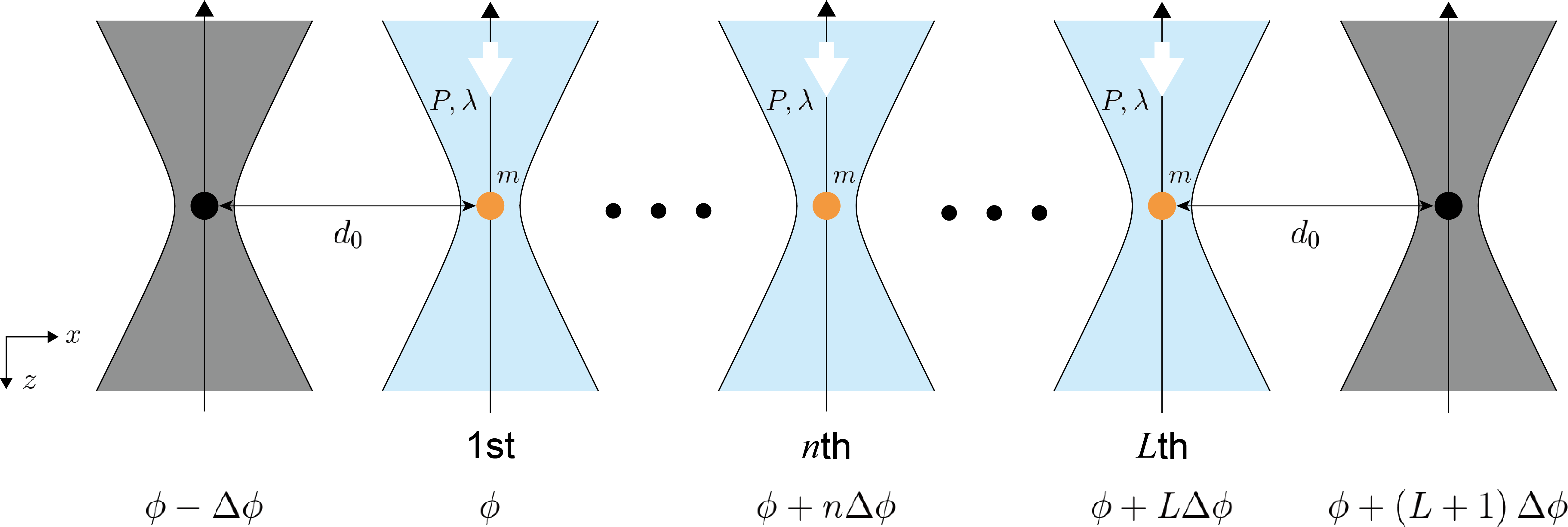}
\caption{\label{figappc1}Schematic figure of the array of the $L$ levitated nanoparticles. The distance between the nearest-neighbor particles is $d_0$, and the mass of all the particles is $m$. The trapping lasers have the power $P$ and the wavelength $\lambda$. We set the phase of the $n$th trapping laser in the focal plane to be $\phi+n\Delta\phi$. At both boundaries of the system, we arrange the two levitated nanoparticles frozen in the motion in all the directions. The trapping lasers at the left and right boundaries have the optical phase $\phi-\Delta\phi$ and $\phi+L\Delta\phi$ in the focal plane, respectively.}
\end{figure*}
We investigate the array of a finite number of levitated nanoparticles as shown in Fig.~\ref{figappc1}. We assume that this system includes only the interaction between the nearest-neighbor particles. Furthermore, at both boundaries of the system, we arrange the deeply trapped levitated nanoparticles, of which the motion in all the directions is frozen, and we set the optical phase of the trapping laser in the focal plane at the left and right boundaries to be $\phi-\Delta\phi$ and $\phi+\left(L+1\right)\Delta\phi$, respectively. In this case, these particles are coupled with the system so that they impose the fixed end boundary conditions on the system.

%
% -------------------------------------------------------------------------------------------------------------------------------------------------------------------------------------
%

\subsection{\label{secC-1}Eigenvalue}
We first show a way to calculate the eigenvalues of the system described by
\begin{eqnarray}
m\ddot{z}_n+m\gamma\dot{z}_n&=&-\left(m\Omega^2+2K\right)z_n \nonumber\\
&&+\left(K+\bar{K}\right)z_{n-l}+\left(K-\bar{K}\right)z_{n+l} \nonumber\\
\label{eqappc1}
\end{eqnarray}
with the fixed boundary condition given by $z_0=z_{L+1}=0$. In this equation, the coupling constants $K$ and $\bar{K}$ are given by Eq.~(\ref{eqappa22}). In the following, we suppose $\left|K\right|\neq\left|\bar{K}\right|$. By assuming $z_n=\psi_ne^{i\omega t}$, Eq.~(\ref{eqappc1}) is rewritten into
\begin{eqnarray}
&&\frac{1}{m}\left[\left(K-\bar{K}\right)\psi_{n+l}\left(K+\bar{K}\right)\psi_{n-l}\right] \nonumber\\
&&+\left(\omega^2-i\gamma\omega-\Omega^2-\frac{2}{m}K\right)\psi_n=0.
\label{eqappc2}
\end{eqnarray}
From a general theory of a difference equation, we can take
\begin{equation}
\psi_n=\sum_{j=1}^2\left(\beta_j\right)^n\phi^{\left(j\right)}
\label{eqappc3}
\end{equation}
as an ansatz of Eq.~(\ref{eqappc2}). Here, $\beta_j\left(=\beta\right)$ is the solution of the characteristic equation given by
\begin{eqnarray}
&&\frac{1}{m}\left[\left(K-\bar{K}\right)\beta+\left(K+\bar{K}\right)\beta^{-1}\right] \nonumber\\
&&+\left(\omega^2-i\gamma\omega-\Omega^2-\frac{2}{m}K\right)=0.
\label{eqappc4}
\end{eqnarray}
We note that Eq.~(\ref{eqappc4}) is a quadratic equation for $\beta$. The boundary conditions, $\psi_0=\psi_{L+1}=0$, tell us the condition that the combination coefficients $\phi^{\left(1\right)}$ and $\phi^{\left(2\right)}$ take nonzero values, and it is written as
\begin{equation}
\left(\frac{\beta_1}{\beta_2}\right)^{L+1}=1.
\label{eqappc5}
\end{equation}
Then, we can obtain the explicit form of $\beta_1$ and $\beta_2$ from Eqs.~(\ref{eqappc4}) and (\ref{eqappc5}). When $\left(K+\bar{K}\right)\left(K-\bar{K}\right)>0$, the Vieta's formula of Eq.~(\ref{eqappc4}) gives
\begin{equation}
\beta_1=re^{i\theta_l},\beta_2=re^{-i\theta_l},
\label{eqappc6}
\end{equation}
where
\begin{equation}
r_+=\sqrt{\frac{K+\bar{K}}{K-\bar{K}}},
\label{eqappc7}
\end{equation}
and
\begin{equation}
\theta_l=\frac{\pi l}{N+1}~\left(l=1,\dots,N\right).
\label{eqappc8}
\end{equation}
Hence, the eigenvalues of the system can be calculated as
\begin{equation}
\omega_{l,\pm}^>=\frac{i}{2}\gamma\pm\sqrt{\Omega^2+\frac{2}{m}\left(K-\sqrt{K^2-\bar{K}^2}\cos\theta_l\right)-\frac{\gamma^2}{4}}.
\label{eqappc9}
\end{equation}
Similarly, when $\left(K+\bar{K}\right)\left(K-\bar{K}\right)<0$, we obtain the form of $\beta_1$ and $\beta_2$ as follows:
\begin{equation}
\beta_1=-ir^\prime e^{i\theta_l},\beta_2=-ir^\prime e^{-i\theta_l}.
\label{eqappc10}
\end{equation}
Here,
\begin{equation}
r_-=\sqrt{\left|\frac{K+\bar{K}}{K-\bar{K}}\right|},
\label{eqappc11}
\end{equation}
and $\theta_l$ is given by Eq.~(\ref{eqappc8}). The eigenvalue of the system in this case is written as
\begin{equation}
\omega_{l,\pm}^<=\frac{i}{2}\gamma\pm\sqrt{\Omega^2+\frac{2}{m}\left(K-i\sqrt{\left|K^2-\bar{K}^2\right|}\cos\theta_l\right)-\frac{\gamma^2}{4}}.
\label{eqappc12}
\end{equation}

%
% -------------------------------------------------------------------------------------------------------------------------------------------------------------------------------------
%

\subsection{\label{secC-2}Finite-size effect}
\begin{figure}[]
\includegraphics[width=8.5cm]{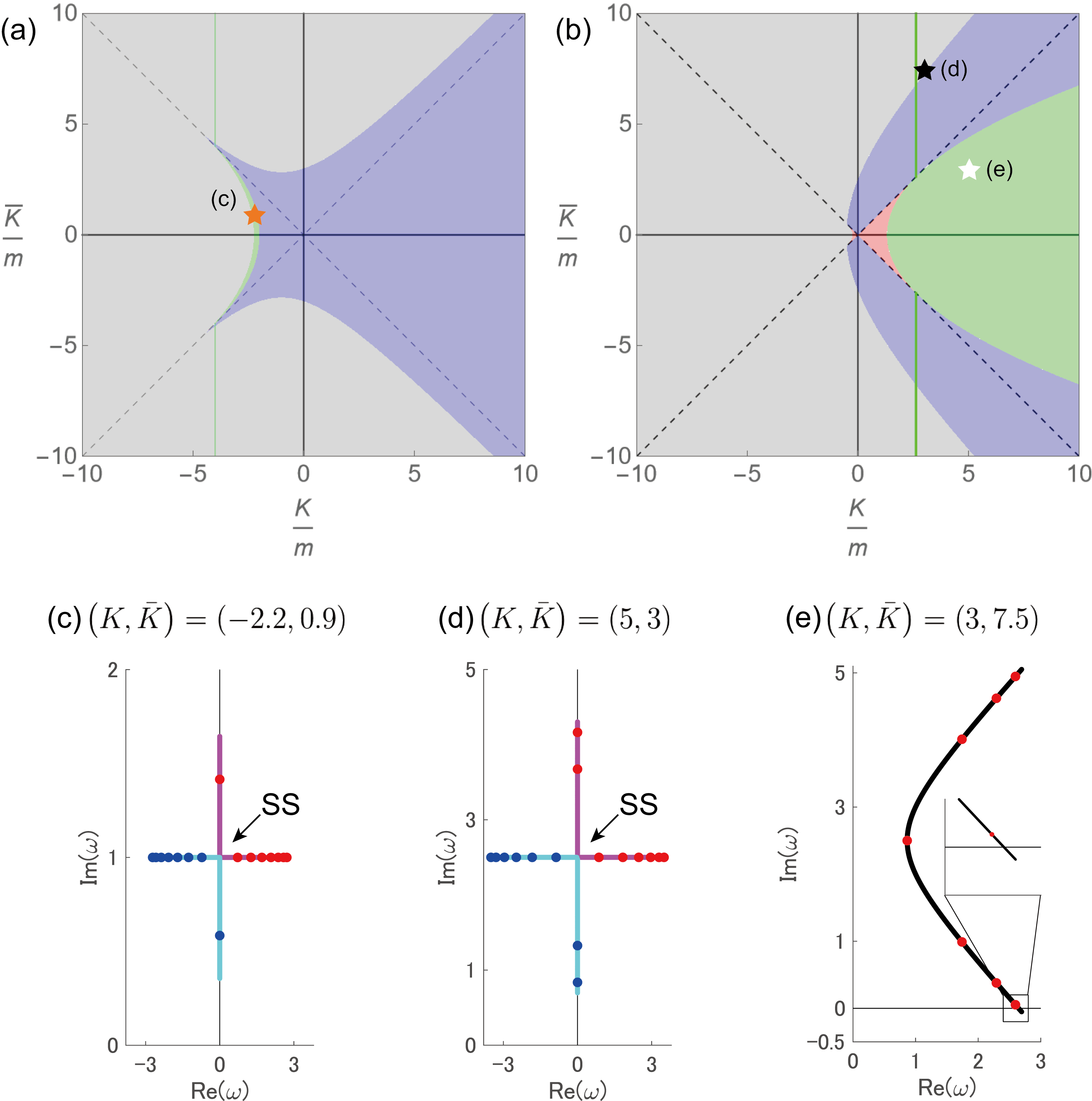}
\caption{\label{figappc2}Dynamical phase diagram, continuum bands, and eigenvalues of the levitated nanoparticle array. (a,b) Dynamical phase diagrams with $\Omega=3$ and $\gamma=2$ and $\Omega=1$ and $\gamma=2$. respectively. Panel (b) is the same as Fig.~\ref{fig2}(a). The blue, green, red, and gray-shaded regions indicate the underdamped, critical, overdamped, and dynamically unstable phase, respectively. The spectral singularity (SS) appears in the green-shaded region and on the green lines. (c) and (d) Continuum bands $\tilde{\omega}_+$ (magenta) and $\tilde{\omega}_-$ (cyan), and eigenvalues $\omega_{l,+}^>$ (red) and $\omega_{l,-}^>$ (blue). (e) Continuum band $\tilde{\omega}_+$ (black), and eigenvalues $\omega_{l,+}^<$ (red). We set $L=8$ in panels (c--e) and choose the values of $\left(K,\bar{K}\right)$ at the orange star for panel (c) and at the white and black stars for panels (d) and (e), respectively.}
\end{figure}
In our work, we investigate the levitated nanoparticle array in the thermodynamic limit, $L\rightarrow\infty$, and obtain the dynamical phase diagram and the continuum bands of the levitated nanoparticle array. We show the dynamical phase diagram with $\gamma<2\Omega$ and $\gamma>2\Omega$ in Figs.~\ref{figappc2}(a) and \ref{figappc2}(b), respectively. We note that the critical phase extends over only a narrow region in Fig.~\ref{figappc2}(a), while it clearly possesses a broad region in Fig.~\ref{figappc2}(b). We here discuss how the finite-size effects can affect the band structures in the critical phase and the boundaries between the underdamped and dynamically unstable phases.

To do so, we recall that the continuum bands of the infinite-size system is given by
\begin{equation}
\tilde{\omega}_\pm=\frac{i}{2}\gamma+\sqrt{\Omega^2+\frac{2}{m}\left(K-\sqrt{K^2-\bar{K}^2}\cos\theta\right)-\frac{\gamma^2}{4}}.
\label{eqappc13}
\end{equation}
First, we consider the parameters indicated by the orange star in Fig.~\ref{figappc2}(a) and show the corresponding continuum bands $\tilde{\omega}_\pm$ and the eigenvalues $\omega_{l,\pm}^>$ in Fig.~\ref{figappc2}(c). One can see that there is no degeneracy between $\omega_{l,+}^>$ and $\omega_{l,-}^>$ in a strict sense, simply because $\theta_l$ takes only the discrete values determined by Eq.~(\ref{eqappc9}). Nevertheless, we emphasize that the eigenvectors around the spectral singularity, where $\tilde{\omega}_+$ touches $\tilde{\omega}_-$, still exhibit the strong nonorthogonality which  is a hallmark of the non-Hermitian degeneracy. This result is the same as in the case of the parameter indicated by the white star in Fig.~\ref{figappc2}(b). Indeed, the eigenvalues in Fig.~\ref{figappc2}(d) are similar to those in Fig.~\ref{figappc2}(c). Next, we consider the parameters indicated by the white star in Fig.~\ref{figappc2}(a) and show the corresponding continuum band $\tilde{\omega}_+$ and the eigenvalues $\omega_{l,+}^<$ in Fig.~\ref{figappc2}(c). It is found that the finite-size system does not exhibit the dynamical instability because the imaginary parts of all the discrete eigenvalues become positive. In this sense, the finite-size effect can slightly modify the boundary between the dynamically unstable phase and the other phases. One can also infer from Fig.~\ref{figappc2}(c) that a larger system size would be favorable to observe the dynamical instability. Nevertheless, we emphasize that the phase diagram of a finite-size system still remains qualitatively the same as in the result obtained in the thermodynamic limit.

%
% -------------------------------------------------------------------------------------------------------------------------------------------------------------------------------------
%

\section{\label{secD}Periodic boundary condition}
\begin{figure}[]
\includegraphics[width=6cm]{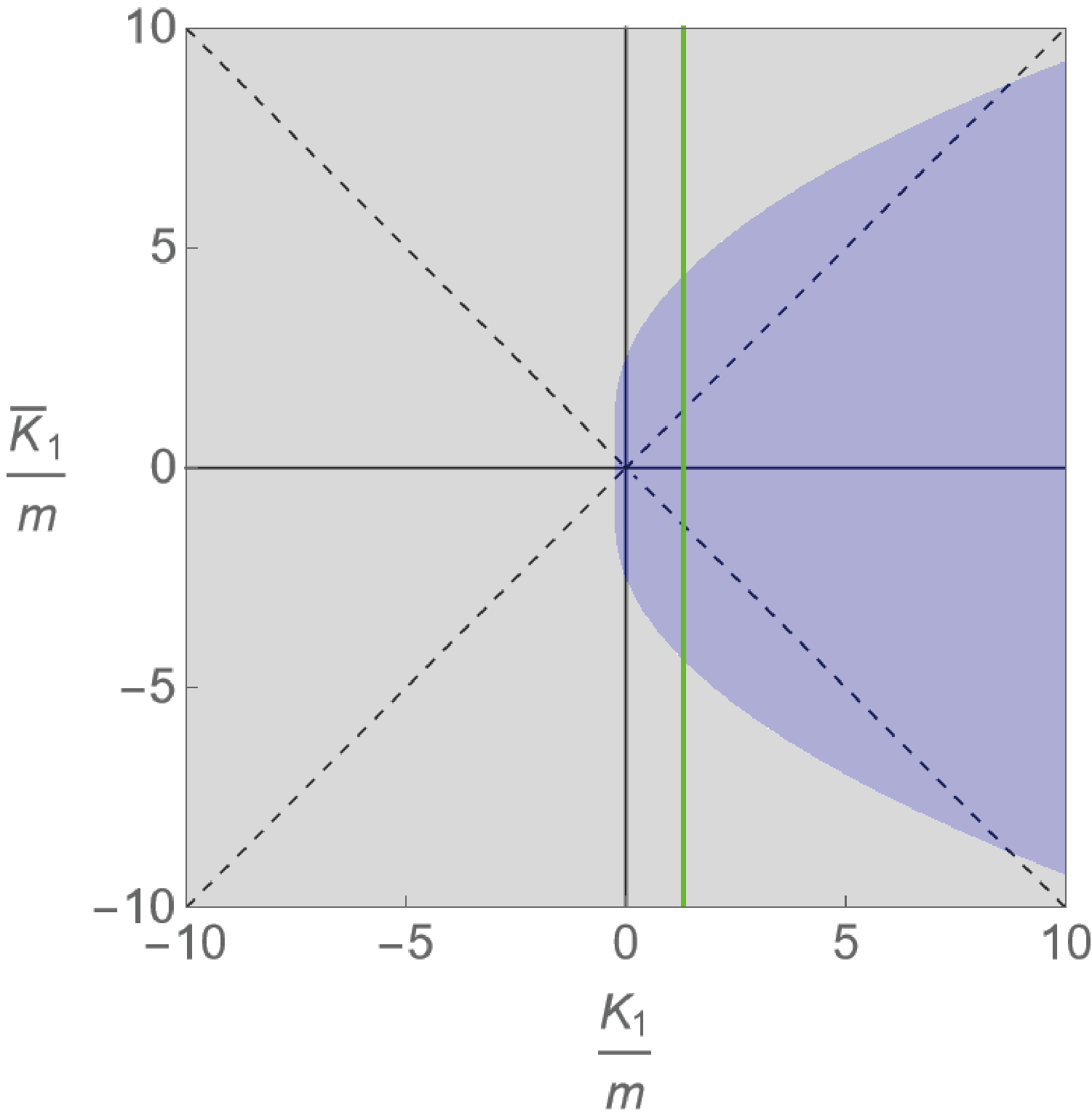}
\caption{\label{figappd}Dynamical phase diagram of the levitated nanoparticle array obtained from the conventional Bloch band theory. The blue and gray-shaded regions are the underdamped and dynamically unstable phases, respectively. The spectral singularity appears on the green line. We set the parameters to be $\Omega=1$ and $\gamma=5$.}
\end{figure}
We consider the levitated nanoparticle array described by Eq.~(\ref{eqappa14}) with $N=1$ and study the dynamical phase diagram under periodic boundary conditions to compare the result obtained in Fig.~\ref{fig2}(a). We note that the characteristic equation of the system reads
\begin{eqnarray}
&&\frac{1}{m}\left[\left(K_1-\bar{K}_1\right)\beta+\left(K_1+\bar{K}_1\right)\beta^{-1}\right], \nonumber\\
&&+\left(\omega^2-i\gamma\omega-\Omega^2-\frac{2}{m}K_1\right)=0.
\label{eqappd1}
\end{eqnarray}
After replacing $\beta$ by $e^{ik}$ for the real Bloch wavenumber $k$, we can obtain
\begin{equation}
\omega_\pm^\prime=\frac{i}{2}\gamma\pm\sqrt{\Omega^2-\frac{\gamma^2}{4}+\frac{2}{m}K_1\left(1+\cos k\right)-\frac{2i}{m}\bar{K}_1\sin k},
\label{eqappd2}
\end{equation}
which reproduces the eigenvalues of the system with periodic boundary conditions. Let us suppose that $\gamma>2\Omega$ for the sake of concreteness. The condition for the appearance of the spectral singularity can be then written as
\begin{equation}
\Omega^2-\frac{\gamma^2}{4}+\frac{4}{m}K_1=0,
\label{eqappd3}
\end{equation}
since it appears when $k=0$. Remarkably, this indicates that the critical phase obtained in Fig.~\ref{fig2}(a) disappears in the system with periodic boundary conditions. Said differently, the appearance of the spectral singularity requires the fine-tuning of the system parameters. Indeed, one can see from the dynamical phase diagram shown in Fig.~\ref{figappd} that only the underdamped and dynamically unstable phases appear in this case. Thus, modifying open boundaries to periodic boundaries drastically changes the behavior of the levitated nanoparticle array.

%
% -------------------------------------------------------------------------------------------------------------------------------------------------------------------------------------
%

\section{\label{secE}Experimental feasibility}
We investigate the dynamical phase diagram of the levitated nanoparticle array with $K/m,\bar{K}/m\in\left[-10\Omega^2,10\Omega^2\right]$ and $\gamma=5\Omega$ in our work. We here discuss the experimental values to access the assumed parameter region in our setup. We note that the previous experiment realizing the coupled two levitated nanoparticles has achieved $K/m,\bar{K}/m\in\left[-0.1\Omega^2,0.1\Omega^2\right]$ and $\gamma\simeq0.03\Omega$~\cite{Rieser2022}, where the power and wavelength of the trapping laser, and the gas pressure are set to be $P=400~{\rm mW}$, $\lambda=1064~{\rm nm}$, and $P_{\rm gas}=1.5~{\rm mbar}$, respectively. Furthermore, the previous work has utilized silica nanoparticles, the radius, the polarizability, and the mass of which are set to be $r=105~{\rm nm}$, $\alpha=3.48\times10^{-32}~{F\cdot m^2}$, and $m=1.07\times10^{-17}~{\rm kg}$, respectively.

First, we explain a possible way to amplify the dipole-dipole couplings $K/m,\bar{K}/m$ compared to $\Omega^2$. To this end, we consider the ration between $G/m$ and $\Omega^2$, which can be obtained from Eqs.~(\ref{eqappa8}) and (\ref{eqappa20}) as follows:
\begin{equation}
\frac{G}{m\Omega^2}=\frac{1}{\pi\varepsilon_0}\left(\frac{k_0^2w_0^2}{2}-1\right)^2\alpha k_0^3.
\label{eqS43}
\end{equation}
It is important that one can achieve $K/m,\bar{K}/m\in\left[-10\Omega^2,10\Omega^2\right]$ by using the trapping laser with the half wavelength and making the radius of the particle $2.5$ times larger than the one used in the previous experiment; we recall that the larger particle results in higher polarizability. Thus, we assume that the wavelength of the trapping laser is set to be $\lambda=532~{\rm nm}$, and the radius, the polarizability and the mass of the particles are set to be $r=260~{\rm nm}$, $\alpha=5.28\times10^{-31}~{\rm F\cdot m^2}$, and $m=1.62\times10^{-16}~{\rm kg}$, respectively, and the laser power is the same as in the previous experiment.

We expect that our model captures qualitative features of the motion of the levitated nanoparticle even if its size is comparable to the wavelength of the trapping laser, because the dominant contribution to the forces which the particle feels still originates from the dipole moments in the particle. Meanwhile, to develop a quantitatively accurate theory, it should be necessary to modify the description to go beyond the dipolar approximation. Specifically, one should take into account the contributions from the higher-order moments (e.g., quadrupole moments).

Second, we explain how to realize the friction coefficient $\gamma=5\Omega$. We note that, in the low-vacuum regime, the damping rate of a levitated nanoparticle is proportional to the square of the radius of the particle and the gas pressure \cite{Millen2020}. With the above parameter set, $\Omega$ approximately becomes twice larger than the value in the previous experiment, and it is possible to realize $\gamma=5\Omega$ by setting the gas pressure to be $P_{\rm gas}=80~{\rm mbar}$. Thus, we expect that the levitated nanoparticle array with the above experimental value set should allow one to realize the parameters assumed in our manuscript and to observe the predicted dynamical phases accordingly.

%
% -------------------------------------------------------------------------------------------------------------------------------------------------------------------------------------
%

%\nocite{*}

\providecommand{\noopsort}[1]{}\providecommand{\singleletter}[1]{#1}%
\end{document}